\documentclass[twocolumn, usenames,dvipsnames]{aastex631}
\usepackage{amsmath, amssymb, xspace, xcolor}
\usepackage[flushleft]{threeparttable}

\newcommand{\p}{\prime}
\newcommand{\mbeq}{\overset{!}{=}}
\newcommand{\mrm}{\mathrm}

\shorttitle{Subphotospheric emission from short gamma-ray bursts}
\shortauthors{Rudolph, Tamborra \& Gottlieb}

\begin{document}

\preprint{INT-PUB-23-038}

\title{Subphotospheric  emission from short gamma-ray bursts: Protons  mold the multi-messenger signals} 

\email{annika.lena.rudolph@nbi.ku.dk}

\author[0000-0003-2040-788X]{Annika Rudolph}
\affil{Niels Bohr International Academy and DARK, 
Niels Bohr Institute, University of Copenhagen, 
Blegdamsvej 17, 
2100, Copenhagen, Denmark}

\author[0000-0001-7449-104X]{Irene Tamborra}
\affil{Niels Bohr International Academy and DARK, 
Niels Bohr Institute, University of Copenhagen, 
Blegdamsvej 17,
2100, Copenhagen, Denmark}

\author[0000-0003-3115-2456]{Ore Gottlieb}
\affil{Center for Interdisciplinary Exploration \& Research in Astrophysics (CIERA), Physics \& Astronomy, Northwestern University, Evanston, IL 60202, USA}
\affil{Center for Computational Astrophysics, Flatiron Institute, New York, NY 10010, USA}
\affiliation{Department of Physics and Columbia Astrophysics Laboratory, Columbia University, Pupin Hall, New York, NY 10027, USA}

\begin{abstract}
The origin of the observed Band-like photon spectrum in short gamma-ray bursts (sGRBs) is a long-standing mystery. We carry out the first general relativistic magnetohydrodynamic simulation of a sGRB jet with initial magnetization $\sigma_0 = 150$ in dynamical ejecta from a binary merger. From this simulation, we identify regions along the jet of efficient energy dissipation due to magnetic reconnection and collisionless sub-shocks. Taking into account electron and proton acceleration processes, we solve for the first time the coupled transport equations for photons, electrons, protons, neutrinos, and intermediate particles species up to close to the photosphere (i.e., up to $1 \times 10^{12}$~cm), accounting for all relevant radiative and cooling processes. We find that the subphotospheric multi-messenger signals carry strong signatures of the hadronic interactions and their resulting particle cascades. Importantly, the spectral energy distribution of photons is significantly distorted with respect to the Wien one, commonly assumed below the photosphere.  Our findings suggest that the bulk of the non-thermal photon spectrum observed in sGRBs can stem from hadronic processes, occurring below the photosphere and previously neglected, with an accompanying energy flux of neutrinos peaking in the GeV energy range.  
\end{abstract}


\section{Introduction} \label{sec:intro}

The conjecture that binary neutron star mergers can power collimated, ultra-relativistic jets~\citep{Eichler:1989ve,Paczynski1991,Nakar:2007yr,Baiotti:2016qnr,Kulkarni:2005jw}, i.e.~short gamma-ray bursts (sGRBs), has been confirmed by the association of  GRB~170817A to the gravitational wave event GW~170817~\citep{LIGOScientific:2017zic,LIGOScientific:2017ync,Mooley:2017enz, Goldstein:2017mmi,Savchenko:2017ffs}.
In the attempt to explain observations and gain much needed insight into the physics of sGRBs, general-relativistic magneto-hydrodynamic (GR-MHD) simulations have been developed with various degrees of sophistication~\citep{Baiotti:2016qnr}. Only  recently multi-dimensional GR-MHD simulations have been able to launch the jet successfully~\citep{Ruiz2016,Ruiz2020,Christie2019,Ciolfi2019,Ciolfi2020,Fernandez2019,Mosta:2020hlh,Nathanail:2020hkx, Gottlieb:2022sis,Gottlieb:2023unified,AguileraMiret2023,Combi:2023yav,Kiuchi:2023obe,Kiuchi:2022nin,Most2023}. Once formed, the jet pierces  through the dynamical ejecta, eventually breaking out~\citep{Gottlieb:2022sis,Mosta:2020hlh,Pavan:2022loy}. The interactions between the jet, the  disk winds, the cocoon, and the magnetic field profile shape the jet structure and composition~\citep{Gottlieb:2020structure,Gottlieb2021:structure,Gottlieb:2022sis,Kiuchi:2023obe,Nativi:2021qzr}, as well as the related particle acceleration and non-thermal emission \citep[e.g.,][]{Lazzati2018,Mooley:2017enz}.

The observed non-thermal nature of sGRB electromagnetic spectra \citep{Ghirlanda:2003ei, Burgess:2017nam, Poolakkil:2021jpc} hints towards efficient particle acceleration taking place in these jets \citep[see e.g.][for the related theoretical background]{Rees:1994nw,Spitkovsky:2008fi}.
However, the physical processes driving particle acceleration in the sGRB prompt phase and the related gamma-ray emission are unsettled. 
Specifically, sGRB observations are difficult to reconcile with optically thin synchrotron models~\citep[][see however also \cite{Burgess:2018dhc}]{Meszaros:2000cc,Ajello:2019zki,Li:2019yik,Ryde:2022lze}. 
An alternative is provided by photospheric models of (modified) thermal spectra released as the outflow becomes optically thin~\citep[see e.g.][]{Ramirez-Ruiz:2005det, Peer:2016mqn, Beloborodov:2017use}.
Such models  may be especially interesting for sGRBs, which generally have harder low-energy spectral index compared to long GRBs~\citep{Poolakkil:2021jpc,Acuner:2017ecd, Dereli-Begue:2020qqx}.

In order to obtain gamma-ray spectra compatible with observations, dissipative effects below the photosphere are required. 
Among the typically invoked processes are
synchrotron and inverse Compton radiation from a non-thermal lepton population \citep{Peer:2005qoc, Giannios:2007yj, Vurm:2012be, Vurm:2015yfa, Bhattacharya:2019pwc}, radiation-mediated shocks \citep{Bromberg:2011by, Lundman:2020oay, Samuelsson:2022fbl} and proton-neutron ($pn$) collisions induced as neutrons drift with respect to the proton flow in the jet \citep{Beloborodov:2009be, Vurm:2011fq}.
While much work on these scenarios has been carried out relying on analytical outflow models, the effects of dissipative processes on the photon spectra remain to be demonstrated in consistent numerical simulations \citep[e.g.~as performed with a Monte-Carlo approach in][]{Ito2015,Ito:2021asl, Ito:2023yhh,Parsotan2018}. In addition, below the photosphere, the photon spectrum could be further modified by lepto-hadronic and hadronic processes~\citep{Kelner:2006tc,Kelner:2008ke}, but these effects have been largely neglected.

If either dissipative processes co-accelerate baryons contained in the jet or heating occurs due to $pn$ collisions, production of high-energy neutrinos is also expected. 
Such sub-photospheric neutrino signals were previously calculated based on simple models \citep{Razzaque:2003uv,Ando:2005xi,Tamborra:2015fzv,Murase:2008sp, Wang:2008zm, Gao:2012ay, Asano:2013jea, Murase:2013hh, Xiao:2017blv}. More recently,  relying on GR-MHD simulations of the jet, neutrino signals at lower energies ($\lesssim 10^5$~GeV)  were predicted both for collapsars \citep{Guarini:2022hry} and sGRBs \citep{Gottlieb:2021pzr}. In fact, because of the strong mixing with the cocoon (obviously not taken into account in na\"{\i}ve  models), the jet is  loaded with baryons, and neutrino production is hindered in the deepest jet regions~\citep{Guarini:2022hry,Gottlieb:2021pzr}. Such recent developments call for advanced modeling of the particle production jointly with the jet evolution in order to forecast the expected multi-messenger signals reliably.

\cite{Gottlieb:2021pzr,Guarini:2022hry}, however, 
focused on the jet region up to several $10^{10}$~cm, thus well below the photosphere (and before breakout from the dynamical ejecta in \citet{Gottlieb:2021pzr}). At these radii, a thermal (Planck or Wien) spectrum was assumed~\citep{Begue:2014kxa,Chhotray:2015lva,Gottlieb:2019aae}. However, as the outflow approaches the photosphere, non-thermal processes may leave imprints both on the photon and neutrino distributions. Hence, it becomes necessary to self-consistently calculate the coupled evolution of photons, protons, leptons and intermediate species. Such non-thermal processes may also have a non-negligible (yet unknown) impact on the observable signals.

This \textit{Letter} explores the multi-messenger signals produced in sGRBs up to close to the photosphere, including leptonic, hadronic as well as lepto-hadronic processes, relying on the benchmark GR-MHD sGRB jet simulation 
introduced in \S~\ref{sec:BNS_sim}. The regions of efficient energy dissipation  along the jet are outlined in \S~\ref{sec:acc_sites}. After introducing the coupled transport equations for all particles  in \S~\ref{sec:mm_methods}, 
we present our results on the  temporal evolution of the spectral energy distributions of photons and neutrinos. A discussion on our findings follows in \S~\ref{sec:discussion}, and we conclude in \S~~\ref{sec:conclusion}. Additional characteristic jet quantities  are provided in Appendix~\ref{appendix:jet}, while details on the numerical evolution of the particle distributions are outlined in Appendix~\ref{appendix:radiative}.

\section{Short gamma-ray burst jet model}
\label{sec:BNS_sim}
\begin{figure*}[t]
    \centering
    \includegraphics[width = 1.0\textwidth] {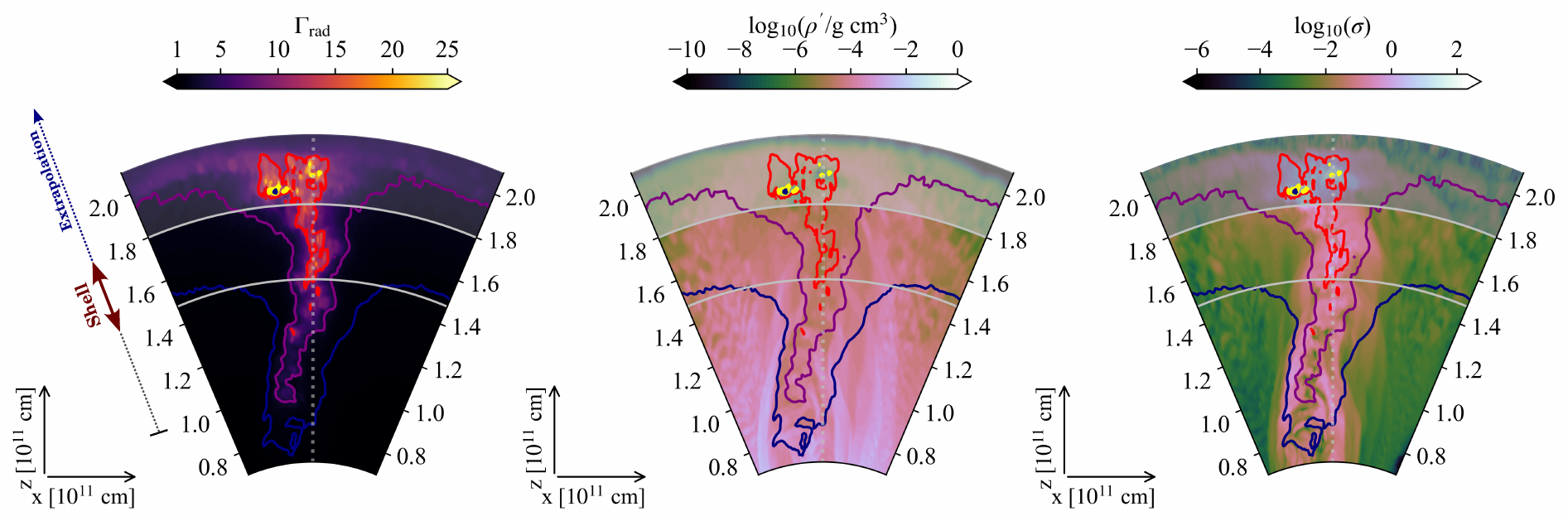}\\
    \vspace{1cm}
    \includegraphics[width = 1.0 \textwidth]{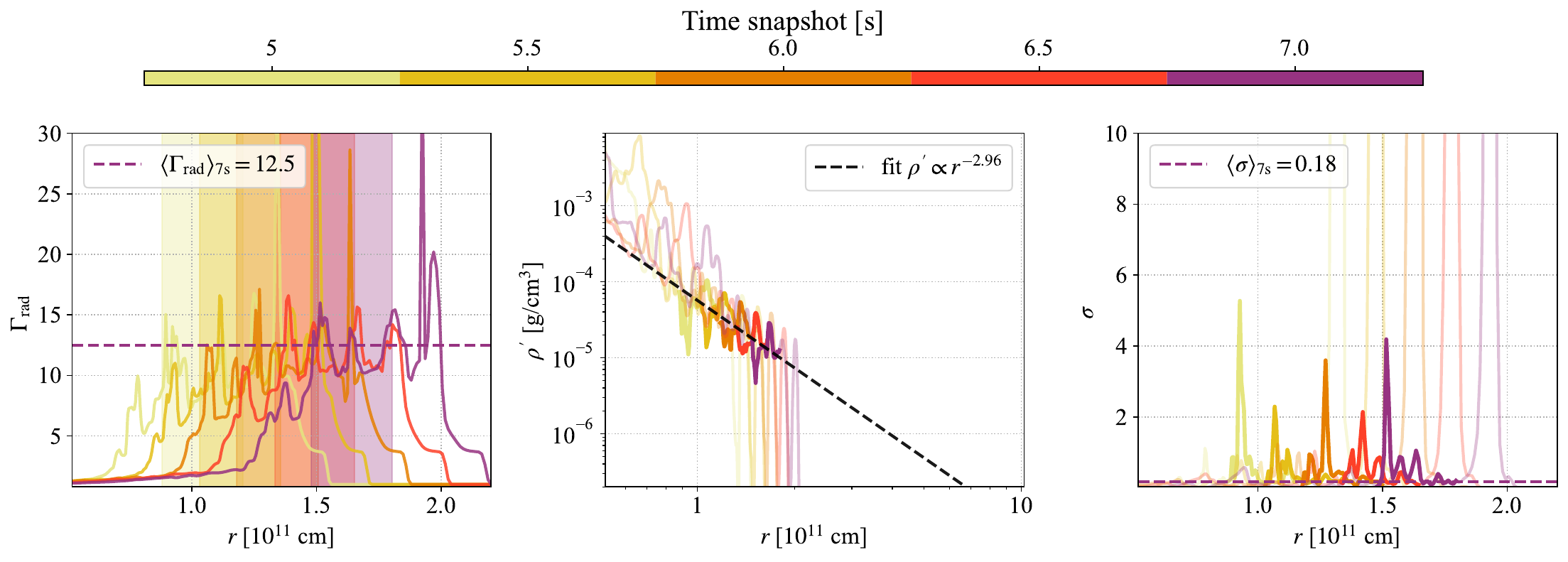}
    \caption{{\it{Top:}} Charateristic properties of our benchmark jet simulation at $7$~s and in the $x$--$z$ plane. From left to right we show the radial component of the Lorentz factor $\Gamma_\mathrm{rad}$, the logarithmic comoving mass density $\log_\mrm{10}(\rho^\prime)$, and the logarithmic magnetization $\log_\mrm{10}(\sigma)$. In order to highlight  the location of the relativistic jet, we plot here the jet region with  viewing angle  $ - 0.4~\mrm{rad} < \theta < 0.4~\mrm{rad}$; the dotted white line marks  $\theta =0$ to guide the eye.  The blue, purple, red, and yellow isocontour lines  correspond to the radial Lorentz factor $\Gamma_\mathrm{rad}$ equal to $1.5$, $3$, $10$, and $25$. At $7$~s, the relativistic jet sits around $1.4$--$2 \times 10^{11}$~cm and it is surrounded by a mildly relativistic cocoon, whose comoving mass density (magnetization) is larger (smaller) than that of the jet.
    The jet simulation inputs from the shaded region at $r \gtrsim 1.8 \times 10^{11}$~cm for the $7$~s snapshot are not considered  in our investigation of the particle acceleration sites.
    Instead, an extrapolation procedure based on a comoving shell located between $1.48$--$1.8 \times 10^{11}$~cm  for the $7$~s snapshot  is adopted  (see main text for details and region delimited by the white solid lines in the top panels); we then extend such extrapolation up to $\simeq 10^{12}$~cm, which is  slightly below the photosphere, i.e.~beyond the jet evolution computed through the GR-MHD simulation.
    We model the acceleration and particle production  following the evolution of the comoving shell, moving out from $1 \times 10^{11}$~cm, as sketched in the left panel.
    {\it{Bottom:}} Evolution of $\Gamma_\mrm{rad}$, $\rho^\prime$ and $\sigma$ for $\theta = \phi = 0$ (thus, along the dotted line in the upper panel) for  $5.0$, $5.5$, $6.0$, $6.5$, and $7.0$~s. In the left plot, we indicate the average Lorentz factor of the last snapshot $\langle \Gamma_\mrm{rad} \rangle_\mrm{7s}$ with a dashed purple line. The shaded band marks  the jet region considered in our multi-messenger emission modeling, which in the two right plots corresponds to the thick, non-transparent lines. In the middle bottom plot of $\rho^\prime $, we show the power-law extrapolation of the density as a dashed black line, while in the right plot of $\sigma$ we indicate the average magnetization at $7$~s within the jet region as dashed purple line (see main text for details). 
    }
    \label{fig:bns_simulation}
\end{figure*}
We perform a GR-MHD simulation of a black hole (BH)-powered jet in  homologously expanding merger ejecta using the GPU-accelerated code \textsc{h-amr} \citep{Liska2022}. We reproduce the fiducial simulation of \citet{Gottlieb:2022sis}, $ {\rm \alpha3d5} $, with the only differences between our and their simulation being that here the jet is launched with a high magnetization of $ \sigma_0 = 150 $ (leading to an asymptotic Lorentz factor of $\mathcal{O}($10-20$)$ compatible with those expected in sGRB jets \citep[e.g.][]{Zou:2017wtv}) and evolves for longer times and larger distances. The setup assumes that the binary merger remnant is a meta-stable neutron star that lasts for $ t_d = 0.5 $~s before collapsing into a BH with mass $ M_{\rm BH} = 3 M_\odot $ and dimensionless spin $ a =0.9375 $. During the meta-stable NS lifetime, a torus with mass $ M_t = 0.2M_\odot $ and a characteristic gas-to-magnetic pressure ratio of $1000$ forms around the compact object. Additionally, the merger debris reaches homologous expansion between velocities $ v_{\rm min} = 0.05 c $ and $ v_{\rm max} = 0.25 c $. Thus, the simulation is initiated with such torus around the BH,  with homologous ejecta with mass $ M_{\rm{ej}} = 0.05\,M_\odot $ (inspired by GW~170817, see e.g.~\cite{Kasen:2017sxr}), and azimuthally-symmetric baryon density profile:
	\begin{equation}\label{eq:density}
		\rho(v_{\rm min}t_d<r<v_{\rm max} t_d,\theta) = \rho_0r^{-3}\left(0.1+{\rm sin}^2\theta\right)\ ;
	\end{equation}
here $\theta$ is the polar angle and $\rho_0$ is determined by requiring that $ M_{\rm{ej}} = 0.05\,M_\odot $. 

We employ an ideal equation of state with a relativistic adiabatic index of $4/3$, within the full simulation setup provided in \cite{Gottlieb:2022sis}.
We follow the jet evolution from its launch by the post-merger BH for $7$~s until its head reaches $ \sim 2 \times 10^{11}$~cm.

The upper panels of Fig.~\ref{fig:bns_simulation} display isocontours of the radial Lorentz factor $\Gamma_\mrm{rad}$, the comoving mass density $\rho^\prime$, and the magnetization $\sigma$, from left to right respectively, extracted at $7$~s for representative purposes, in the $x$--$z$-plane (thus $\phi = 0$; $\phi$ being the azimuthal angle) for $7 \times 10^{10} \, \mrm{cm} < r < 2.1 \times 10^{11} \, \mrm{cm}$, and viewing angle $ - 0.4~\rm{rad} < \theta < 0.4~\rm{rad} $~\footnote{Hereafter,  we denote the quantities in the source and plasma comoving frame as $X$ and $X^\prime$, respectively.}. 
From the profile of $\Gamma_\mrm{rad}$  (left panel), one can  identify the relativistic jet between $1.4$--$2.0 \times 10^{11}$~cm,  roughly propagating along the $z$-axis, and a mildly relativistic cocoon surrounding the jet. The jet extension corresponds to an engine activity time of about $2$~s, roughly consistent with the typical duration of sGRBs.  It is important to note that the outflow is still optically thick at $7$~s with  optical depth $\tau \gtrsim 10^4$ below $2.1 \times 10^{11}$~cm along the $z$--axis (see Fig.~\ref{fig:appendix_jet_parameters1} in Appendix~\ref{appendix:jet}).
The high-density structures surrounding the jet stem from the cocoon (light pink regions in the middle panel of Fig.~\ref{fig:bns_simulation}). In contrast, the relativistic  jet has a lower density. The magnetization, $\sigma= B^{\prime 2}/4 \pi \rho^\prime c^2$ (where $B^\prime$ is the comoving magnetic field), in the right panel of Fig.~\ref{fig:bns_simulation} follows the inverse pattern: the largest $\sigma$ is visible in regions of low $\rho^\prime$.

In order to investigate the efficiency of particle acceleration up to the photosphere (located at $r \gtrsim 10^{12}$~cm) and related multi-messenger emission, we focus on the region with $10^{11} \, \mrm{cm}\lesssim  r \lesssim 10^{12}$~cm. Within this radial range, we follow the comoving evolution of a shell representing the relativistic jet, that e.g.~at $7$~s is  located between $1.48 \times 10^{11} \, \mrm{cm}\lesssim  r \lesssim 1.8 \times 10^{11}$~cm 
(see the top panel of Fig.~\ref{fig:bns_simulation}), 
where the radial Lorentz factor $\Gamma_\mrm{rad} \gtrsim 10$. At the viewing direction $\theta = \phi = 0$, the jet extension in radius is maximal, while the internal energy reaches its highest values. In this sense, it may dominate the overall observed emission and be representative of the jet region.
The jet region with $r \gtrsim 1.8 \times 10^{11}$~cm is shaded  in the top panel of Fig.~\ref{fig:bns_simulation}, as the simulation here is subject to numerical artifacts that plague the evolution of the magnetization.
Consequently, this region is ignored in our post-processing of the jet simulation to compute the multi-messenger emission. 
Since the full evolution up to the photosphere is not captured by the GR-MHD simulation, we rely on extrapolations obtained as follows.

At each radius, the shell is characterized by its Lorentz factor $\Gamma$, magnetization $\sigma$, internal energy density $u^\prime$, and rest mass density $\rho^\prime$. In order to extrapolate the behavior of these quantities, we select the snapshots taken at $[5.0, 5.5, 6.0, 6.5, 7.0]$~s (see the bottom panels of Fig.~\ref{fig:bns_simulation} and Appendix~\ref{appendix:jet}); note that we choose to begin the computation of particle emission at $5$~s  when the jet characteristic quantities almost reached asymptotic values, approaching the end of acceleration and adiabatic expansion, the photon spectral energy distribution being thermal earlier. 
Both the radial Lorentz factor as well as the magnetization change marginally between the last snapshots; thus, we assume constant $\Gamma \equiv \langle \Gamma_\mrm{rad} \rangle_{7\rm{s}} = 12.5$ and $\sigma \equiv \langle \sigma \rangle_{7\rm{s}} = 0.18$ (here, the label $7$~s refers to the considered jet simulation snapshot). 
For  $u^\prime$ and $\rho^\prime$, which due to expansion of the system clearly decay with time, we instead extrapolate the time evolution from the jet simulation. In order to obtain representative values of these quantities up to the photosphere, we first perform a fit for each quantity of interest, relying on the shaded  cells   displayed in the bottom left panel of Fig.~\ref{fig:bns_simulation}  and then calculate the  average of each quantity. This procedure  yields 
$\rho^\prime (r) = 2.3 \times 10^{28}
 \left({r}/{\rm{cm}}\right)^{-2.96}\, \mathrm{g/cm^3}$ and 
 $u^\prime (r) = 5.6 \times 10^{55} \left({r}/{\rm{cm}}\right)^{-3.56}\, \mathrm{erg/cm^3}$.
 Note that this deviates from the behavior expected for the simple, self-similar expansion of the shell (that would yield $\rho^\prime \propto r^{-2}$). 
 This hints   that the acceleration has not fully ceased  (in fact, we find that the average  specific enthalpy $\langle h^\prime \rangle= \langle 1 + 4 p^\prime/(\rho^\prime c^2) \rangle \gtrsim 2$ along $\theta=\phi=0$ in the considered region).
 Future GR-MHD simulations up to even larger radii are imperative to tackle this problem fully.
These fits  are somewhat dependent on $\theta$, yet  for all viewing angles we find  that  the decay of $\rho^\prime$ is steeper than $r^{-2}$ and  the difference in the power-law indexes for $u^\prime$ and $\rho^\prime$ is between $0.2$ and $0.4$, with $u^\prime$ always decaying faster than $\rho^\prime$. The latter potentially affects the formation of the Wien spectrum stemming from the Planck spectrum and the evolution of the thermal photon peak~\citep{Vurm:2012be, Begue:2014kxa}.

\section{Particle acceleration sites}
\label{sec:acc_sites}
\begin{figure}[htb]
    \centering
    \includegraphics[width = 0.45\textwidth]{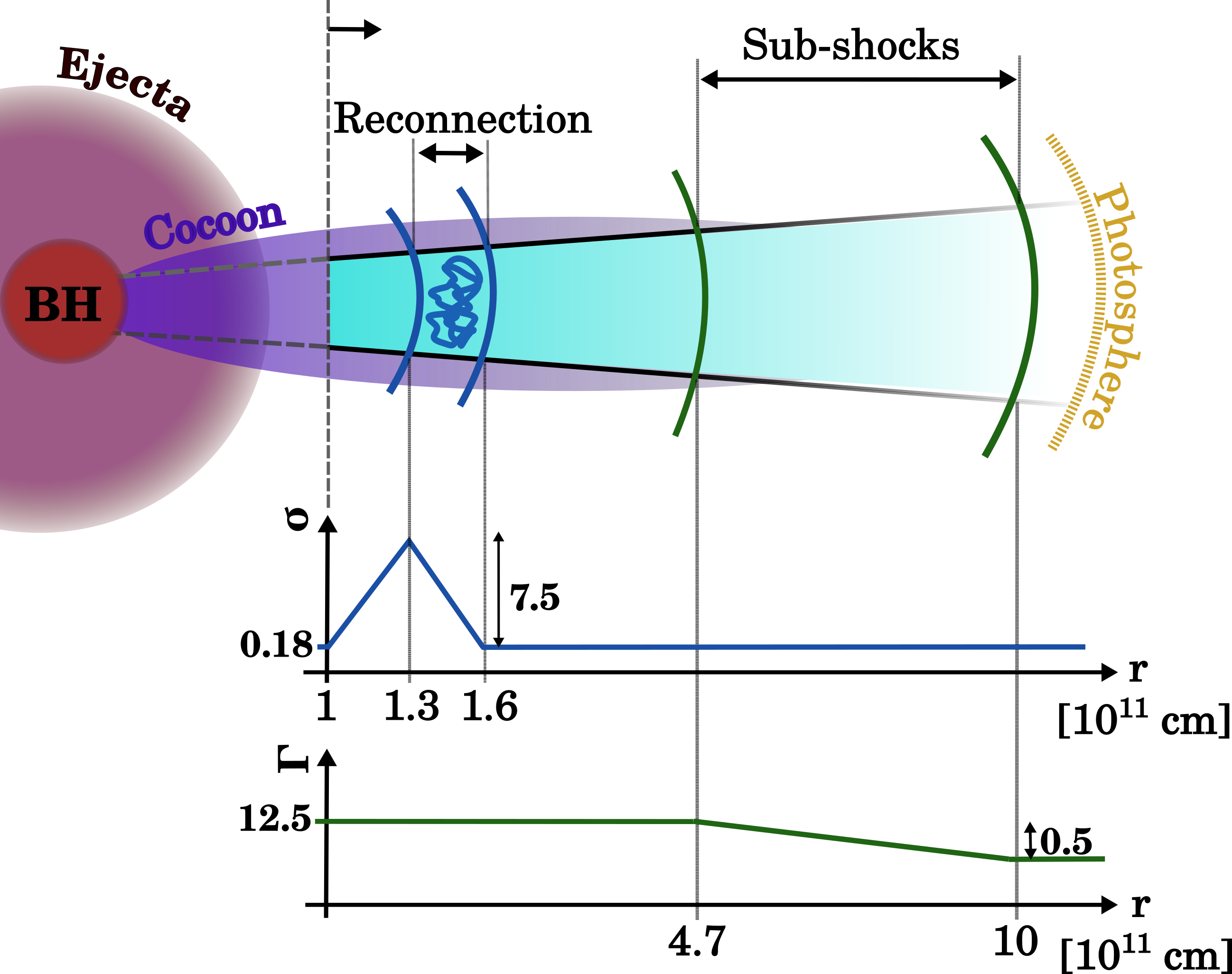}
    \caption{Sketch of the particle acceleration sites in our sGRB jet simulation (not to scale). 
    The central BH (dark red) launches a jet (cyan) which is surrounded by a mildly-relativistic cocoon (dark purple). The jet propagates outwards, breaking out from the  merger ejecta (pink); note that the jet  is not present anymore at smaller radii since it  was active for $\simeq 2$~s--the  sub-relativistic  cocoon expands in the region previously occupied by the jet. At $10^{11}$~cm, it has reached a magnetization $\sigma = 0.18 $ and Lorentz factor $\Gamma = 12.5 $; we neglect the jet region below this radius, since we are interested in processes close to the photosphere potentially shaping the observed spectrum. A local increase in $\sigma$ occurs between $ 1$--$1.3 \times 10^{11}$~cm, while the decay of $\sigma$ enables magnetic reconnection (equivalent to $1$~s  duration). The induced change in magnetization is $\Delta \sigma = 7.5$. Variability in the Lorentz factor of the jet leads to collisionless sub-shocks between $4.7 \times 10^{11}$~cm and $1 \times 10^{12}$~cm; as a consequence, a decrease in the Lorentz factor of $\Delta \Gamma = 0.5 $ occurs. 
    The photosphere at which the jet becomes optically thin is represented through the solid ochre line.}
    \label{fig:sketch}
\end{figure}

We now identify the regions of energy dissipation in the jet and subsequent particle acceleration, focusing on the jet region at $r \gtrsim 10^{11}$~cm and assuming the representative shell evolution introduced above. 
Analyzing our benchmark simulation, we identify the following mechanisms of energy dissipation:
\begin{enumerate}
    \item  Magnetic reconnection. The low magnetization $\sigma = 0.18$ in principle inhibits magnetic reconnection~\citep{Blandford:1977ds,Drenkhahn:2002ug,Drenkhahn:2001ue,Gill:2020oon}. However, the magnetization is locally enhanced up to $\sigma = 7.5 $ in the jet region between $1$--$1.6 \times 10^{11}$~cm and for  the narrow range of $\theta$ covering the relativistic jet, with slight variations in the exact positions/distributions of the peaks. Since it is too expensive to solve the transport equations (see \S~\ref{sec:mm_methods}) while considering the small fluctuations in $\sigma$, we take into account these local maxima of $\sigma$ by assuming  $\left| \mathrm{d} \sigma / \mathrm{d} r \right| =$~const., where  the magnetization builds up between $1$--$1.3 \times 10^{11}$~cm before decaying between $1.3$--$1.6 \times 10^{11}$~cm (see Fig.~\ref{fig:sketch}). Although this simplified profile may not be representative of the complex reality, we do not expect this choice to impact our results as long as the average $\left| \mathrm{d} \sigma / \mathrm{d} r \right| $ and $\sigma$ are similar in both cases. Following the GR-MHD simulation, we do not account for magnetic dissipation above $1.6 \times 10^{11}$~cm. The induced rate of energy dissipation during the decay of the magnetic field is
        \begin{equation}
        \frac{\mathrm{d}u_B^\prime}{\mathrm{d}t^\prime} = \rho^\prime c^2 \beta \Gamma c \frac{\mathrm{d}\sigma}{\mathrm{d} r} \, .
        \end{equation}
        
    \item Internal shocks and collisionless sub-shocks. If the jet Lorentz factor varies on a timescale $t_v$ (length scale $c t_v$), internal shocks take place  at the radius $2 t_v \Gamma^2 c $~\citep{Rees:1994nw}. 
    We identify variability in the  jet Lorentz factor of our benchmark simulation on timescales $t_v \sim 50$--$500$~ms. Internal shocks are thus expected at $4.7 \times 10^{11} \, \mathrm{cm} \lesssim r \lesssim 4.7 \times 10^{12} \, \mathrm{cm} $.
    These shocks convert kinetic into internal energy, which causes a decrease in the Lorentz factor of the plasma. Hence,  the dissipated energy can be estimated through the expected slow-down of the plasma.
    To compute the dissipated energy, we follow a procedure similar to the simple internal shock model \citep[e.g.][]{Barraud:2005wr} based on the interaction between a fast (f) and a slow (s) plasma shell. 
    Within the considered region, the fastest plasma moves with $\Gamma_\mathrm{f} \simeq 16$ at $r_\mathrm{f} = 1.5 \times 10^{11}$~cm, whereas the slowest plasma has $\Gamma_\mathrm{s} \simeq 9$ at $r_\mathrm{s} = 1.5 \times 10^{11} + t_v c$~cm. For the maximal variability timescale of $t_v = 500$~ms, this simple model yields $\Delta \Gamma \simeq 0.5$. Although single shocks may dissipate energy at specific radii, the continuous velocity profile can be expected to yield a smooth dissipation of energy. We emulate this by assuming $\mathrm{d}\Gamma/\mathrm{d}r = \Delta \Gamma / \Delta r_\mrm{IS} =$const. 
    We thus compute the comoving rate of energy dissipation via internal shocks as
        \begin{equation}
            \frac{\mathrm{d}u_{\rm{IS}}^\prime}{\mathrm{d}t^\prime} = \rho^\prime c^2 \beta \Gamma c \frac{\mathrm{d}\Gamma}{\mathrm{d} r} \, .
            \label{eq:du_dt_internalshocks}
        \end{equation}
    When internal shocks are radiation mediated (as expected in our case for the optically thick plasma below the photosphere, $\tau \gtrsim 1$ in the region of $10^{11} \, \mrm{cm} \lesssim r \lesssim 10^{12}\, \mrm{cm}$; see Appendix~\ref{appendix:jet}), particle acceleration is inhibited~\citep{Levinson:2019usn}. However, collisionless sub-shocks that form within the radiation-mediated shock may enable particle acceleration if the plasma is mildly magnetized ($ 0.1 \lesssim \sigma \lesssim 1$) and 
        \begin{equation}
        \chi \equiv  2 u^\prime / 3 \sigma \rho^\prime c^2 \lesssim 2 
        \end{equation}
    holds \citep{Beloborodov:2016jmz}. While, for simplicity, we do not follow numerically the evolution of the shock structure consistently with the spectra of the accelerated particles~\citep{Budnik:2010ru,Beloborodov:2016jmz},  at $4.7 \times 10^{11} \, \mathrm{cm} \lesssim r \lesssim 1 \times 10^{12} \, \mathrm{cm} $, we find that collisionless sub-shocks meet the  criteria introduced above and are therefore an efficient mean of accelerating particles in our jet model;  
    the energy released in  sub-shocks is obtained through Eq.~\ref{eq:du_dt_internalshocks}  (which is effectively an upper limit, for the sub-shocks embedded in the radiation mediated shocks).
    \item MHD turbulence. Acceleration at MHD turbulences (possibly in connection with shocks) has been proposed as an alternative particle acceleration mechanism in GRBs~\citep{Lemoine:2021mtv, Bresci:2023pjx}. Given that it is not expected to  operate efficiently in the collisional plasma below the photosphere, it is omitted in our modeling. 
\end{enumerate}
In synthesis, as sketched Fig.~\ref{fig:sketch}, in  our benchmark simulation, the jet crosses three different regions as it propagates between $10^{11}$--$10^{12}$~cm: 1.~The reconnection region in the range $[1, 1.6] \times 10^{11}$~cm, where a spike in $\sigma$ induces magnetic reconnection; 2.~the expansion region, in the range  $[1.6, 4.7] \times 10^{11}$~cm, where no additional energy dissipation occurs;  3.~the sub-shock region above $4.7 \times 10^{11}$~cm, where particles are accelerated at collisionless sub-shocks.
In Fig.~\ref{fig:sketch},  we further schematically show how $\sigma$ and $\Gamma$ are affected in the dissipation regions. 
Note that the photosphere is located still at slightly larger radii, as can also be inferred from the Compton-$Y$-parameter that equals $\sim 10$ at  $10^{12}$~cm (see Fig.~\ref{fig:appendix_jet_parameters2} in Appendix\ref{appendix:jet}).

\section{Multi-messenger emission}
\label{sec:mm_methods}

We now introduce the transport equations of electrons, positrons, protons, neutrons, pions, muons, neutrinos, and photons, as well as their related particle distributions. We also present our findings on the temporal evolution of the particle spectral distributions.

\subsection{Particle transport equations}
In a purely thermal plasma, electrons and photons may interact through Compton scatterings, as well as through  thermal bremsstrahlung, double Compton, and cyclotron emission or absorption.
In the case of efficient particle acceleration, numerous leptonic and hadronic emission processes may shape the multi-wavelength spectra: First, all charged particles in a magnetic field emit synchrotron radiation and participate in inverse Compton scatterings. 
The plasma may be enriched by energetic electron-positron pairs by means of $\gamma \gamma$-annihilation or photo-hadronic pair production, where the secondary pairs may contribute significantly to the photon spectra through the synchrotron and inverse Compton processes.
Finally, hadronic processes such as proton-proton interactions and photo-pion production yield energetic pions that decay into photons, muons, electrons, and neutrinos. Here, the secondary neutrinos serve as unique probes of  hadronic interactions.

Using the inputs at each $r$ from our jet model, as illustrated in \S~\ref{sec:BNS_sim}, we solve the time-dependent coupled partial differential equations  of (non-thermal) electrons, positrons, protons, neutrons, pions, muons, neutrinos and photons:
\begin{align}
\label{eq:transport}
    \delta_t n(x) = & \frac{1}{A(x)} \delta_x \left( D(x)\delta_x n(x) + a(x) n(x) \right) \nonumber \\ 
    & + \varepsilon(x) - \alpha(x) n(x) \, ,
\end{align}
with  $\delta / \delta_x \equiv \delta_x$,  $n$ being the particle number (or occupation number, for momentum space), and $x$ representing the dimensionless particle energy/momentum.
In our treatment for photons, $x$ represents the momentum and the pre-factor $A(x) = 4 \pi x^2$.
For all other (non-thermal) particle species, $x$ represents the energy and the pre-factor $A(x) = 1$. The remaining factors account for diffusion [$D(x)$], advection/cooling [$a(x)$], source/injection [$\varepsilon(x)$], and sink/escape [$\alpha(x)$]. 

The energy distributions of non-thermal electrons, positrons, protons, neutrons, pions, muons and neutrinos are evolved relying on a modified version of the \textsc{AM$^{3}$} code~\citep{Gao:2016uld}. In addition, we extend the original  code that includes synchrotron, inverse Compton, photo-pion production, Bethe-Heitler pair production, $\gamma \gamma $-annihilation, and adiabatic cooling; our new version of \textsc{AM$^{3}$} also accounts for proton-proton interactions (see Appendix~\ref{appendix:radiative}). 
Secondary particles are treated self-consistently, undergoing the same interactions as primary particles.

The photon evolution is instead computed through a newly developed code. The latter treats the evolution  in momentum space, accounting for Comptonization by a thermal electron population through the Kompaneets equation \citep{1957:Kompaneets}. We extend the Kompaneets equation to include cooling, absorption/emission terms due to thermal emission and absorption processes (bremsstrahlung, cyclotron, and double Compton) as well as the effects of adiabatic expansion. Non-thermal processes are included by offering an interface to \textsc{AM$^3$} ($\gamma \gamma $-annihilation, synchrotron, inverse Compton, and pion decay). The initial spectrum at $r \simeq 10^{11}$~cm
is a Planck black-body spectrum. For the thermal protons, neutrons, and electrons, no separate treatment is necessary. For these instead, at each time step, the temperature and number density are updated, relying on the evolution of the internal and rest mass density of the plasma shell. The assumption of Maxwellian distributions for the thermal populations is justified since the thermal relaxation time due to Coulomb collisions is shorter than the dynamical time, for the radial range under consideration. 
For  details on the considered processes and the numerical treatment, we refer the interested reader to Appendix~\ref{appendix:radiative}.
The interface between the GR-MHD simulation and the radiative calculations is thus implemented on the one hand through the thermal populations (following the evolution of $\rho$ and $u$), and on the other hand through the injection terms for the non-thermal electrons and protons that are a result of the inferred particle acceleration regions and processes.

\subsection{Accelerated particle distributions}
We describe the accelerated  spectrum for the particle species $i$ (i.e., protons or electrons; see  injection term in Eq.~\ref{eq:transport}) as a power-law distribution with exponential cutoffs both at the minimum Lorentz factor $\gamma^\prime_\mrm{min}$ and the maximal Lorentz factor $\gamma^\prime_\mrm{max}$: 
\begin{equation}
\varepsilon (\gamma_i) \equiv \frac{\mrm{d}n^\prime_i}{\mrm{d}\gamma_i^\prime \mrm{d}t^\prime}= N_i \gamma_i^{\prime -p_i} e^{-\gamma^\prime_{i, \mrm{min}}/ \gamma_i^\prime}e^{-\gamma_i^\prime/ \gamma^\prime_{i, \mrm{max}}} \, .
\end{equation}
The  factor $N_i$ is computed by normalizing with respect to the energy dissipation rate $\mrm{d}u^\prime / \mrm{d}t^\p$. 
The splitting of $\mrm{d}u^\prime / \mrm{d}t^\p$ between the different particle populations is parameterized through $\epsilon_{\rm{th}}$ (that defines the fraction of energy transferred to thermal protons, neutrons, and electrons) and $\epsilon_i$ (the fraction of  non-thermal energy transferred to particles of species $i$).
Then, $N_i$ can be calculated through
\begin{equation}
   \epsilon_i \frac{\mrm{d} u^\prime }{\mrm{d}t^\prime } \mbeq \int \mrm{d}\gamma^\prime_i \: \gamma^\prime_i m_i c^2 \frac{\mrm{d}n^\prime_i}{\mrm{d}\gamma^\prime_i\mrm{d}t^\prime} \, .
\end{equation}
Similarly, the thermal plasma is heated by $\epsilon_{\rm{th}} {\mrm{d} u^\prime }/{\mrm{d}t^\prime }$. The latter increases the temperature of the thermal proton, neutron and electron populations. 
The minimum and maximum Lorentz factors, as well as the power-law slopes $p_i$, $\epsilon_{th}$, $\epsilon_i$ depend on the acceleration process at work. As these parameters specify the properties of the primary injected distributions, they are not altered by the radiative processes. In contrast, the cooled comoving distributions may differ significantly; for example, secondary lepton pairs can dominate the electron distribution.

We summarize our assumptions on the factors entering the accelerated particle distributions for magnetic reconnection and collisionless sub-shocks in Table~\ref{tab:acc_particle_distr}. Here, we assume that magnetic reconnection operates similar to electron-ion reconnection in the optically thin regime, as explored e.g.~in \citet{Sironi:2015eoa, Werner:2016fxe, Zhang:2023lvw}. However, even in a collisionless plasma, magnetic reconnection may be slowed down due to the injection of electron-positron pairs \citep{Hakobyan:2018fwg}---further work is required regarding reconnection in collisional plasmas \citep{Uzdensky:2011ka}. 

The maximum particle energy attainable for acceleration at sub-shocks can be computed by balancing the acceleration characteristic timescale with the energy losses (and their corresponding loss rates $t^\prime_\mrm{loss}$). 
For a particle of Lorentz factor $\gamma_i$ and mass $m_i$,  the acceleration rate  is~\citep{Globus:2014fka}
\begin{equation}
    t^{\prime -1}_\mathrm{acc}(\gamma^\prime_i) =  0.1 q_i B^\prime / m_i c \gamma^\prime_i \, , 
\end{equation}
with $q_i$ being  the particle charge.
As for magnetic reconnection, the synchrotron burnoff may be exceeded for acceleration along the electric field, where the maximum energy scales with the magnetization~\citep[see e.g.][]{Cerutti:2013mma, Kagan:2016rze, Kagan:2017rls}, thus $\gamma_\mathrm{e, max}^\prime =  5 \sigma m_\mathrm{p} / m_\mathrm{e}$ and $\gamma_\mathrm{p, max}^\prime = (\gamma_\mrm{e, max}^\prime -1) m_\mrm{e}/ m_\mrm{p} +1 $.

\begin{table}[htb]
    \centering
    \begin{threeparttable}
        \footnotesize
        \caption{Summary of the quantities entering the accelerated particle distributions.}
        \label{tab:acc_particle_distr}
        \begin{tabular}{c  c c }
            \hline
            Quantity & Definition & Reference \\ 
            \multicolumn{3}{c}{\textbf{Magnetic reconnection}} \\ \hline 
             $\gamma_\mathrm{e, min}^\prime $ &$  \gamma_\mrm{e, max}^\prime/40$ & K18\\
            $\gamma_\mathrm{e, max}^\prime $ &$  5 \sigma m_\mathrm{p} / m_\mathrm{e}$ & K18 \\
            $p_\mathrm{e} $ &$   1.9 + 0.7 / \sqrt{\sigma}$ & W18\\
             $\gamma_\mathrm{p, min}^\prime $ &$  (\gamma_\mrm{e, min}^\prime -1) m_\mrm{e}/ m_\mrm{p} +1 $& K18\\
             $\gamma_\mathrm{p, max}^\prime $ &$  (\gamma_\mrm{e, max}^\prime -1) m_\mrm{e}/ m_\mrm{p} +1 $ & K18 \\
            $p_\mathrm{p} $ &$  1.9 + 0.7 / \sqrt{\sigma}$ & W18\\
             $\epsilon_\mathrm{e} $ &$  1/8 (1 + \sqrt{\sigma /10+ \sigma})$ & W18\\ 
             $\epsilon_\mathrm{p} $ &$  1/2 - \epsilon_\mathrm{e} $ & W18\\ 
            $\epsilon_\mathrm{th} $ &$   0.5$  & S15\\ \hline
            \multicolumn{3}{c}{\textbf{Sub-shocks}} \\ \hline
            $\gamma_\mathrm{e, min}^\prime $ &$  2 m_\mathrm{p}/ m_\mathrm{e}$ & C19 \\
             $\gamma_\mathrm{e, max}^\prime $ &$  \sum t^{\prime -1 }_\mathrm {loss}(\gamma^\prime_\mathrm{max}) \mbeq t^{\prime -1}_\mathrm{acc} (\gamma^\prime_\mathrm{max}) $ & \\
             $p_\mathrm{e} $ &$   2.2$ & C19 \\
             $\gamma_\mathrm{p, min}^\prime $ &$  1$ & C19 \\
            $\gamma_\mathrm{p, max}^\prime $ &$  \sum t^{\prime -1 }_\mathrm {loss}(\gamma^\prime_\mathrm{max}) \mbeq t^{\prime -1}_\mathrm{acc} (\gamma^\prime_\mathrm{max}) $ & \\
            $p_\mathrm{p} $ &$  2.0$ & C19 \\
             $\epsilon_\mathrm{e} $ &$  5 \times 10^{-4}$ & C19 \\ 
             $\epsilon_\mathrm{p} $ &$   0.1$ & C19\\ 
             $\epsilon_\mathrm{th} $ &$  1.0 - \epsilon_\mathrm{e} - \epsilon_\mathrm{p}$ & C19 \\ \hline
        \end{tabular}
        \begin{tablenotes}
            \small
            \item
            \textit{Notes:} 1.~We neglect non-thermal proton acceleration in magnetic reconnection for $\sigma \lesssim 10$. In this case, $\epsilon_p {\mrm{d} u^\prime }/{\mrm{d}t^\prime }$  is instead a heating term for the thermal populations.
            2.~Reference abbreviations: K18--\citet{Kagan:2017rls}, W18--\citet{Werner:2016fxe}, S15--\citet{Sironi:2015eoa}, C19--\citet{Crumley:2018kvf}. 
        \end{tablenotes}
    \end{threeparttable}
\end{table}

\subsection{Temporal evolution of the distributions of photons and neutrinos}
\label{sec:results}

\begin{figure*}[hbt]
    \centering
    \includegraphics[width = 1.0 \textwidth]{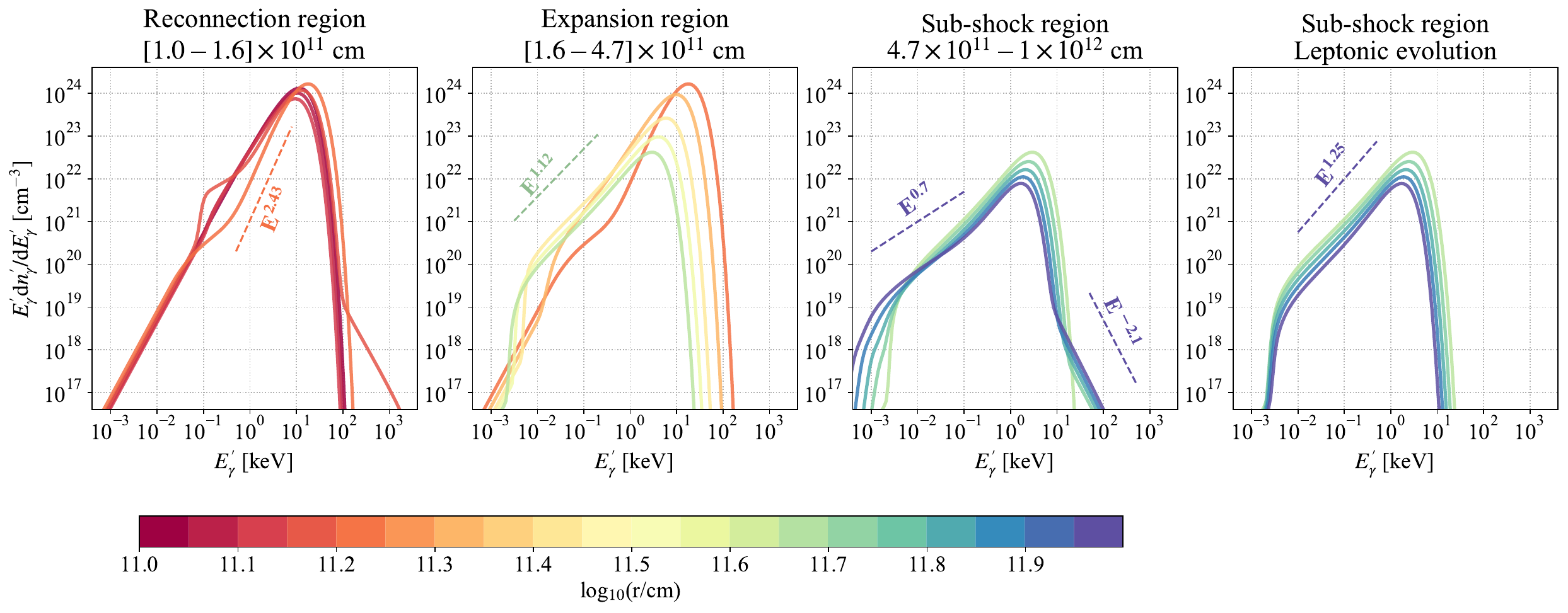}
    \caption{Evolution of the comoving spectral energy distribution of photons with radius, split in three regions (see also Fig.~\ref{fig:sketch}). In the magnetic reconnection region (first panel from the left), the local enhancement of the magnetization enables magnetic reconnection and the subsequent acceleration of electrons. In the expansion region (second panel), no energy dissipation takes place and therefore the photon distribution peak shifts to lower energies, with overall  lower  photon density as a result of the plasma expansion. Finally, in the sub-shock region (third panel), both protons and electrons are accelerated at collisionless sub-shocks, leading to the appearance of a  non-thermal high-energy tail and  a  softening of the spectrum below the distribution peak. In order to highlight the impact of hadronic processes,   the photon distribution obtained without non-thermal protons is also shown in the  sub-shock region  (fourth panel). To highlight  the evolution of the photon distribution, a power-law fit  at the final snapshot of each region is provided.}
    \label{fig:photon_evolution}
\end{figure*}
\begin{figure}[htb]
    \centering
    \includegraphics[width = 0.5\textwidth]{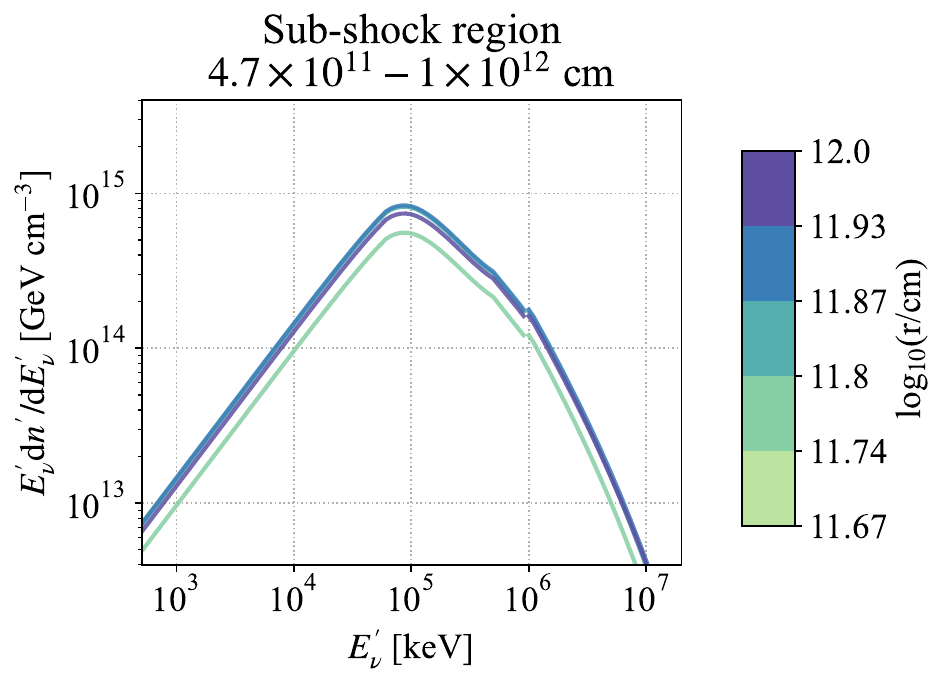}
    \caption{Evolution of the comoving spectral energy distribution of  neutrinos (summed over all flavors). Neutrino production is dominated by proton-proton interactions. We only focus on the third region (sub-shock region), since neutrino production is inefficient in the region of magnetic reconnection. Note that no escape term is included in the differential equation for the neutrinos; and the shown spectra rather represent time-integrated results (corrected for adiabatic expansion). The same color scale as in Fig.~\ref{fig:photon_evolution} is adopted.}
    \label{fig:neutrino_evolution}
\end{figure}

Figures~\ref{fig:photon_evolution} and \ref{fig:neutrino_evolution}  show our findings on the time evolution of the photon and neutrino spectral energy distributions, respectively, between $10^{11}$~cm and $10^{12}$~cm obtained by solving Eq.~\ref{eq:transport} and relying on the evolution of the shell introduced above. In the following, we outline the main features of the particle distributions considering the three jet regions introduced before (see Fig.~\ref{fig:sketch}):
\begin{enumerate}
    \item {Magnetic reconnection region.} During the ramp-up of the magnetization $\sigma$ at $1.0 \times 10^{11} \, \mrm{cm} < r < 1.3 \times 10^{11} \, \mrm{cm}$, the photon distribution evolves as a thermal one; adiabatic expansion and the decreasing thermal electron temperature are responsible for the shift of the distribution peak at lower energies and a decrease in number density. The slope of such a spectrum is $E^\prime n_\gamma (E^\prime_\gamma) \propto E^{\prime 2}_\gamma$.
    
    As magnetic reconnection becomes active for $r \gtrsim 1.3 \times 10^{11} \, \mrm{cm}$, two effects manifest: 1.~A non-thermal photon population is injected. The latter appears as  a non-thermal tail  at $E^\prime_\gamma \gtrsim 10^2$~keV in the initial power-law distribution.
    At lower energies, efficient Comptonization of the synchrotron seed photons induces a softening of the spectrum between $E^\prime_\gamma \sim 10^{-1}$--$1$~keV. At the lowest energies, the plasma is still in thermal equilibrium, its thermal shape  induces a ``bump''-feature at the turnover energy. 2.~Due to the low $\sigma \lesssim 10$, protons are not accelerated above thermal energy. Subsequently, roughly half of the dissipated energy heats the thermal population. 
    Since photons and electrons are still coupled, the peak of the photon thermal spectrum is consequently shifted to higher energies. As the energy dissipation ceases and the electron acceleration stops, the non-thermal signatures directly disappear and a narrow Wien spectrum approximately scaling as $E^{\prime 2.43}_\gamma$ is evident (light orange line in the leftmost panel of Fig.~\ref{fig:photon_evolution}). Due to proton acceleration being inefficient, no neutrino production is expected.
    
    \item {Expansion region.} As the jet expands without further energy dissipation, the plasma cools and dilutes. Hence, the peak of the photon distribution moves to lower energies, and lower photon densities are achieved. The redistribution of photons abundantly present due to the previous dissipation phase yields a spectrum scaling approximately as $E^{\prime 1.12}_\gamma$(in the region where the spectrum is not in thermal equilibrium for $E^\prime_\gamma \gtrsim 10^{-2}$~keV---see the second panel to the left of Fig.~\ref{fig:photon_evolution}). As the plasma moves outwards, the turnover energy at which photons are still in equilibrium shifts to lower energies, broadening the the soft part of the spectrum.

    \item {Sub-shock region.} Protons and electrons accelerated at collisionless sub-shocks lead to the formation of a non-thermal high-energy tail, as well as a strong softening of the spectrum below the distribution peak (with a spectrum scaling approximately as $E^{\prime 0.7}_\gamma$, as displayed in the third panel of  Fig.~\ref{fig:photon_evolution}; the thermal distribution peak is approximately preserved despite the dissipation processes. 
    
    We highlight  the important role of hadronic processes in shaping the photon distribution. In fact, if only electron acceleration is accounted for (rightmost  panel of  Fig.~\ref{fig:photon_evolution}), the photon distribution basically  maintains its former shape. This strong effect of hadronic dissipation is linked to the opacity of the plasma, where a secondary cascade of lepton pairs is injected due to $\gamma \gamma$-annihilation of the very-high-energy pion decay photons as well as through Bethe-Heitler pair production. 
    In the sub-shock region, proton-proton interactions operate efficiently, being the dominant energy loss process for low-energy protons. 
    
    The evolution of the neutrino energy distribution produced by the subsequent decay of energetic pions is depicted in Fig.~\ref{fig:neutrino_evolution}. The relatively low maximum energy of neutrinos is attributed to efficient cooling of protons due to photo-pion and Bethe-Heitler processes. We also find that the neutrino energy distribution is mainly shaped by proton-proton interactions, the proton-photon channel being sub-leading and leading to the production of neutrinos with slightly higher energies on average.   Note that we neglect the contribution of kaons in shaping the neutrino energy distribution.  Kaon decay is expected to affect the neutrino spectrum at high energies~\citep{Ando:2005xi,Asano:2006zzb}. Below the photosphere and for a collapsar jet with $\sigma_0 = 200$, \cite{Guarini:2022hry} showed that kaon decay leads to a bump above $\mathcal{O}(10^3)$~GeV in the neutrino distribution due to the large magnetic field. However, in our case, the low maximum energy of neutrinos and the dominance of proton-proton interactions suggest that the kaon contribution would lead to negligible modifications.   
    
    Note that our simulation includes no escape, but an adiabatic expansion term for neutrinos. In this sense, the plotted energy distributions are  representative of time-integrated ones, although corrected for the increasing comoving volume. Let us highlight two aspects characterizing the temporal evolution of the neutrino distribution. 
    After the first (green) line, the later neutrino spectra are almost identical with their corresponding lines lying on top of each other. This is indicative of an almost steady-state proton population. 
    Second, the peak of the neutrino distribution lies around $10^5$~keV. Note that, if represented as energy flux, this peak is shifted to an energy of $1$~GeV, which for a typical redshift of $ z = 2$ would yield an observed peak energy of $E_\nu = 4.1 $~GeV. This rather low peak energy may be understood by the  low maximum proton energies, where photo-pion reactions potentially producing neutrinos of higher energies are inhibited, similar to the findings of \citet{Guarini:2022hry}.
\end{enumerate}

Overall, the period of magnetic energy dissipation at small radii tends to  increase the number of available photons and yields a softened spectrum in the Wien zone. The cascades initiated by hadronic processes in the sub-shock region soften these spectra even further, in addition to a high-energy non-thermal tail. 

The magnitude of the reported signatures naturally scales with the energy dissipation rate. Further, we focus on specific regions where magnetic reconnection and internal shocks are expected. However, magnetic dissipation and sub-shocks due to variability on shorter timescales, may also affect the spectrum in the expansion region, but are not addressed in this work.

\section{Discussion}
\label{sec:discussion}
The origin of the spectral shape of the gamma-ray spectrum observed for (s)GRBs  remains unclear~\citep{Nakar:2007yr,Baiotti:2016qnr}.  Optically thin synchrotron spectra tend to yield spectra that are too soft/wide, while (re-processed) thermal spectra are rather too hard or narrow.
On the other hand, the photon spectral energy distribution, shown in Fig.~\ref{fig:photon_evolution} at $r \sim 10^{12}$~cm, scales as $\mrm{d} n^\prime/ \mrm{d} E^\prime_\gamma \propto E^{\prime -0.3}_\gamma$ below the peak.
This behavior is compatible with the power-law slopes observed for a fraction of sGRBs, although in most cases the spectral index at low energies is still lower \citep{Burgess:2017nam, Poolakkil:2021jpc}. This could be resolved through enhanced energy dissipation, such as stronger shocks due to larger differences in Lorentz factors, considering more shocks or an extended magnetic reconnection region. Further, softer spectra can be obtained if the evolution of $u$ and $\rho$ is such that the energy released by sub-shocks in comparison to the internal energy density is larger than in the current scenario. 

It is important to stress that, if hadronic processes are not accounted for, the photon distribution is significantly harder with $\mrm{d} n^\prime/ \mrm{d} E^\prime_\gamma \propto E^{\prime 0.25}_\gamma$ (see  Fig.~\ref{fig:photon_evolution}), which is potentially difficult to reconcile with sGRB gamma-ray spectra. Note that similar to the neutrino spectra, photon spectra are mostly impacted by $pp$-interactions instead of photo-pion production. At high energies, we find that synchrotron emission of secondary non-thermal electrons results in a high-energy tail of the spectrum scaling as $\mrm{d} n^\prime/ \mrm{d} E^\prime_\gamma \propto E^{\prime - 3.1}_\gamma$, which is steeper than the average observed slope---although for sGRBs a low high-energy spectral index has also been reported \citep{Acuner:2017ecd, Poolakkil:2021jpc}. Again, this signature is only visible if hadronic processes are included. Being dominated by secondary pairs, the high-energy slope differs from the one expected for an electron population with $p_e \sim 2.1$--$2.2$.
Finally, the comoving peak energy of photons at $\mathcal{O}(1$~keV$)$, obtained in Fig.~\ref{fig:photon_evolution} and a relic of the peak of the thermal distribution at $10^{11}$~cm, would require a high(er) jet Lorentz factor in order to be compatible with typical sGRB peak energies of $\mathcal{O}(10^2$--$10^3$~keV$)$~\citep{Burgess:2017nam, Poolakkil:2021jpc}. 
Here we point out that noticeable relics of the thermal spectrum are still present around the peak in our final photon distribution;  stronger energy dissipation, potentially also taking place beyond the photosphere, may be required to wash these features out \citep[see also][]{Vurm:2011fq, Vurm:2015yfa}. 

Dominated by the products of proton-proton interactions, the neutrino distribution [$E_\nu^{\prime} \mrm{d}n /\mrm{d} E_\nu$] obtained in our simulation and shown in Fig.~\ref{fig:neutrino_evolution}, peaks at $E_\nu^\prime \simeq 0.02$~GeV.
The corresponding energy flux  [$E_\nu^{\prime 2} \mrm{d}n /\mrm{d} E_\nu$] peaks  at $E_\nu^\prime \simeq 1$~GeV, 
equivalent  to an observed energy of $E_\nu = 4.1 $~GeV (for a typical redshift of $z = 2$) as a consequence of the low Lorentz factor $\Gamma = 12.5$. This finding is compatible with the results presented for collapsars in \citet{Guarini:2022hry} and  with the ones of \citet{Gottlieb:2021pzr} obtained for sGRBs, where the production of neutrinos at collisionless sub-shocks at $r \sim 10^9$~cm was considered and a maximal neutrino energy of $10^3$~GeV  obtained. Relying on the findings of  \citet{Guarini:2022hry,Gottlieb:2021pzr}, we expect that the low-energy tail of the  neutrino distribution might be boosted by less efficient neutrino production occurring  at $r \lesssim 10^{11}$~cm. However, our results are in contrast with the ones previously reported in e.g.~\citet{Razzaque:2004yv,Ando:2005xi,Tamborra:2015fzv} and \citet{Murase:2013ffa} for low-luminosity GRBs, forecasting that neutrinos  with  TeV--PeV energy could be produced in choked jets harbored in collapsars because of collisionless internal shocks or collimation shocks (the latter are  unlikely to be efficient in our case because of the large optical depth of the jet). The neutrino spectra from the jet could be complemented by neutrino emission from other components, see e.g.~ \citep{Decoene:2019eux,Kimura:2018vvz}. However, we  expect these to be sub-dominant along the line of sight of the jet.

The detection of this low-energy neutrino flux may be challenged by the atmospheric background in this energy range, however timing information could favor its detection in  Hyper-Kamiokande and IceCube Deep Core~\citep{Guarini:2022hry}.

Future work should be addressed to simulate the long-term evolution of sGRB jets and the related neutrino and radiation transport, reaching  up to optical depth $\tau \simeq \mathcal{O}(0.1)$. 
In this context, the still ongoing acceleration  of the plasma should be accounted for. Note, however, that as long as the magnetization in the sub-shock region 
supports
collisionless subshocks, we do not expect a major impact on our main conclusions.
The evolution of the shock structure should be also modeled  consistently with the  spectra of the accelerated particles.
Moreover, growing evidence points towards a major impact of the magnetic field profile, its amplification mechanism, as well as a non-trivial coupling between energy deposition by neutrino annihilation and the magnetic field  in the jet launching and its dynamics~\citep{Christie2019,Ciolfi:2020cpf,Mosta:2020hlh,Gottlieb:2022sis,Gottlieb:2023unified,Gottlieb:2023est,Gottlieb:2023vuf,Hayashi2022,Hayashi2023,Combi:2023yav,Kiuchi:2022nin, Kiuchi:2023obe}; all these factors may further affect the sites of energy dissipation as well as the associated multi-messenger emission. 

Observations of sGRBs suggest a large variety in the temporal evolution of their signals. This may be due to a different distribution of the  energy dissipation regions, as well as a different jet magnetization, and magnetic field configurations. We intend to investigate the impact of the  jet properties on the multi-messenger observables in a companion paper~\citep{Rudolph:2023}.  Along the same vein, one may also consider different shells propagating outwards and representing a non-homogeneous jet region, in addition to taking into account the dependence on the viewing angle/angular jet structure.

\section{Conclusions}
\label{sec:conclusion}
In this \textit{Letter}, we explore the multi-messenger signatures of sub-photospheric energy dissipation in sGRBs. We carry out a state-of-the art GR-MHD simulation of a binary neutron star merger aftermath, where we follow the propagation of a relativistic jet with initial magnetization $\sigma_0 = 150$ in massive ejecta for $7$~s. For the first time,  we simulate the coupled evolution of non-thermal and thermal particles in a plasma shell representative of the jet region that propagates between $10^{11}$ and $10^{12}$~cm (the  photosphere being located at a slightly larger radius), including all relevant thermal and non-thermal processes. The evolution of the characteristic parameters of the plasma shell (such as the internal energy, rest mass density, Lorentz factor, and magnetization) have been extrapolated from our GR-MHD  simulation up to $10^{12}$~cm,  the Lorentz factor and magnetization having already reached their asymptotic values at $10^{11}$~cm. 

In our model, we  identify a region of magnetic energy dissipation at $[1, 1.6] \times 10^{11}$~cm, where  $\sigma$ is locally enhanced by $7.5$ enabling magnetic reconnection. While acceleration of protons above non-thermal energies is inhibited (hence, neutrinos cannot be produced), radiation of non-thermal leptons introduces a weak non-thermal high-energy tail and a low-energy feature in the spectral distribution of photons.
In the region between $[1.6, 4.7] \times 10^{11}$~cm, the plasma expansion cools the thermal populations, and  the photons are redistributed at lower  energies, with a consequent softening of the photon distribution  below the thermal peak.  

We identify variability in the Lorentz factor, leading to internal shocks  between $[4.7 \times 10^{11}, 10^{12}]$~cm. Although particle acceleration is inefficient at internal shocks in a radiation dominated environment, protons and electrons can be efficiently accelerated at collisionless sub-shocks because of the mild magnetization in the region $[4.7 \times 10^{11},  10^{12}]$~cm considered in our model. 
The resulting non-thermal radiation from the primary accelerated particles and the induced secondary cascade significantly alters the spectrum with low- and high-energy asymptotic distributions scaling as $n^\prime_\gamma (E_\gamma^\prime) \propto E_\gamma^{\prime -0.3}$ and $n^\prime_\gamma (E_\gamma^\prime) \propto E_\gamma^{\prime -3.1}$, while the distribution peak derives from the thermal one at $10^{11}$~cm. The main modifications of the spectral energy distribution of photons in this region are due to hadronic processes,  highlighting the need to include such processes and the resulting particle cascades when modeling sub-photospheric GRB spectral energy distributions. The obtained asymptotic slopes are compatible with the one observed in a fraction of sGRBs. Further energy dissipation not accounted for in the present scenario (both below and above the photosphere) may yield photon distributions closer to the typical observed spectra.  
The spectral peak of the photon energy distribution is softened 
 during the expansion to $\sim 1$--$10$~keV, which may require a larger Lorentz factor in order to reach the typical observed peak energy of sGRBs.
Neutrinos are also produced in the sub-shock region, where efficient proton-proton interactions, Bethe-Heitler induced cooling as well as  sub-dominant photo-pion production lead to a distribution peaking in the sub-GeV to GeV regime. 

We conclude that photon spectra compatible with sGRB observations may be produced from radiation of protons and leptons accelerated below or close to the photosphere. Initiating an effective cascade, hadronic processes are crucial for transitioning from a Wien distribution to a true non-thermal one. Signatures of hadronic processes may be naturally carried by neutrinos as well.
Future work extending our analysis to and above the photosphere, as well as including contributions from different viewing angles would be desirable to further confirm our findings.

\section*{Acknowledgments}
We are grateful to Eli Waxman for insightful discussions and to the anonymous referee for constructive feedback. This project has received funding from the Villum Foundation (Project No.~37358), the Carlsberg Foundation (CF18-0183), and the Deutsche Forschungsgemeinschaft through Sonderforschungsbereich SFB 1258 ``Neutrinos and Dark Matter in Astro- and Particle Physics'' (NDM). IT also thanks the  Institute for Nuclear Theory at the University of Washington for its
hospitality and the Department of Energy for partial support during the final stages of this work. OG is supported by Flatiron Research and CIERA Fellowships. OG also acknowledges support by Fermi Cycle 14 Guest Investigator program 80NSSC22K0031, and NSF grant
AST-2107839.
An award of computer time was provided by the ASCR Leadership Computing Challenge (ALCC), Innovative and Novel Computational Impact on Theory and Experiment (INCITE), and OLCF Director's Discretionary Allocation programs under award PHY129. This research used resources of the National Energy Research Scientific Computing Center, a DOE Office of Science User Facility supported by the Office of Science of the U.S. Department of Energy under Contract No. DE-AC02-05CH11231 using NERSC award ALCC-ERCAP0022634.

\bibliography{references}
\bibliographystyle{aasjournal}



\appendix
In this Appendix, we provide additional details on the jet characteristic properties as well as on the numerical implementation of the radiative processes and the related evolution of the particle distributions.

\section{Jet characteristic properties}
\label{appendix:jet}
\begin{figure}[htb]
    \centering
    \includegraphics[width = 0.5\textwidth]{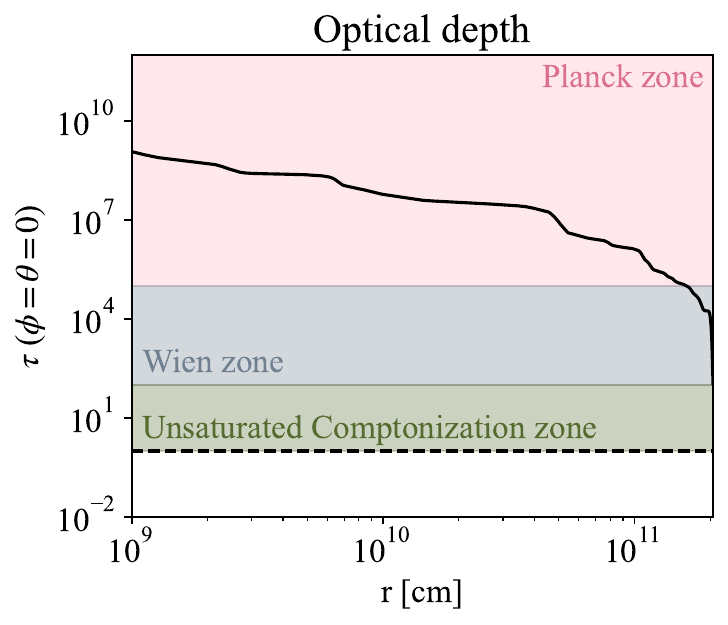}
    \caption{Optical depth along the $z$--axis (corresponding to $\theta = \phi = 0$ ) from our sGRB jet simulation at $7$~s. We mark the Planck zone with $\tau \gtrsim 10^{5}$, the Wien zone with $10^2 \lesssim \tau \lesssim 10^{5}$, and the Comptonization zone with $\tau \lesssim 10^2 $. 
    }
    \label{fig:appendix_jet_parameters1}
\end{figure}

\begin{figure}[htb]
    \centering
    \includegraphics[width = 0.7\textwidth]{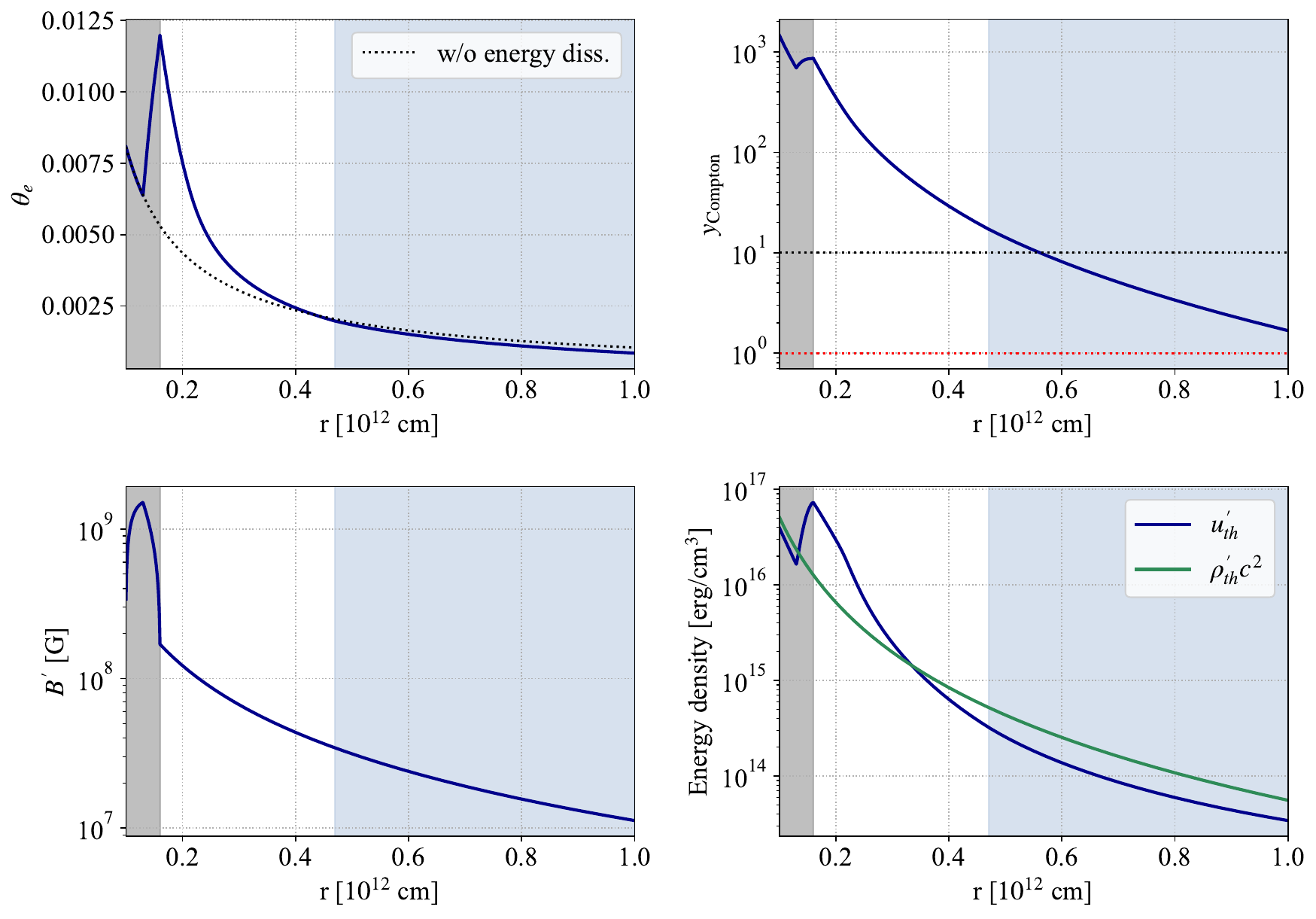}  
    \caption{Evolution of the jet characteristic quantities within the  shell considered for particle acceleration, as it propagates outwards from $10^{11}$~cm to $10^{12}$~cm. The upper  panels show the dimensionless electron thermal electron temperature (calculated from the electron temperature $T_e$ as $\theta_e = k_B T_e/m_e c^2$) on the left and the Compton-$y$ parameter with a [black, red] dotted horizontal line indicating where [$y_\mrm{Compton} = 10$, $y_\mrm{Compton} = 1$] on the right. The lower panels show the comoving magnetic field $B^\prime$ (left) as well as the energy density and rest mass density of the thermal population (composed of protons, neutrons and electrons, on the right). In all plots we shade the magnetic reconnection region in grey, and the sub-shock region in light blue.}
    \label{fig:appendix_jet_parameters2}
\end{figure}

Additional information on the jet characteristic properties is displayed in Figs.~\ref{fig:appendix_jet_parameters1} and ~\ref{fig:appendix_jet_parameters2}. Figure~\ref{fig:appendix_jet_parameters1} shows the optical depth on-axis (thus for $\theta = \phi = 0)$ at $7$~s as a function of radius. One can see that, at the end of the simulation time, the jet is still in the optically thick regime of the Wien zone. Hence, any non-thermal signatures can be expected to be thermalized and the processes shaping the observed spectra would rather occur at radii larger than the simulated ones; for this reason,  the jet evolution up to the photosphere needs to be extrapolated. 

Figure~\ref{fig:appendix_jet_parameters2}  shows the evolution of the characteristic quantities within the considered shell as it expands in radius. In the top left panel, the evolution of the dimensionless electron temperature $\theta_e = k_B T_e/m_e c^2$ is shown. For comparison, we  include the analytical evolution expected for a purely thermal scenario with a black-body photon field, without taking into account  effects of energy dissipation. Clearly, the dissipation in the magnetic reconnection region strongly increases the temperature of the thermal populations. The impact of the magnetic reconnection region is further clearly visible in the evolution of the magnetic field, displayed in the bottom left panel. 

In the upper right panel of Fig.~\ref{fig:appendix_jet_parameters2}, we show the Compton-$y$ parameter calculated as \citep{Vurm:2011fq}
\begin{equation}
    y_\mathrm{Compton} = 4 \theta_e \sigma_T n_e t^\prime_\mrm{dyn} \, ,
\end{equation}
where $ t^\prime_\mrm{dyn} = r / 2 \beta \Gamma c$.
From the plot we conclude that at the final radius of $10^{12}$~cm, efficient Comptonization is still guaranteed. 
Finally, the bottom right panel shows the energy density and rest mass density of the thermal populations. Except for the region where magnetic reconnection enhances the energy density, the rest mass density exceeds the internal energy one.

\section{Details on the numerical implementation of the radiative processes and particle distribution evolution}
\label{appendix:radiative}
In this Appendix, we provide details about the radiative processes and their numerical treatment. In order to capture all particle populations, we employ three different approaches as described below.  
First, we evolve the temperature and number densities of the thermal populations. We have verified that, for the parameters and jet radial range of interest, neutrons, protons, and electrons remain in equilibrium (i.e., the Coulomb relaxation time is  much smaller than the dynamical time of the system). The density of each particle species is proportional to  $\rho^\prime (r)$ of the plasma shell (with the electron-to-baryon ratio $Y_e \simeq 0.5$, see e.g.~\cite{Just:2022flt}). The evolution of the thermal temperature $ \theta_e$ can be obtained by equating 
\begin{equation}
 \frac{\mathrm{d} u^\prime }{\mathrm{d} t^\prime} = \frac{3 m_e c^2 (1 + 1/Y_e) \rho^\prime}{2 m_p}  \frac{\mathrm{d} \theta_e }{\mathrm{d} t^\prime} + \frac{3 m_e c^2}{2 m_p}\theta_e (1+1/Y_e) \frac{\mathrm{d} \rho^\prime }{\mathrm{d} t^\prime} + \left[\frac{\mathrm{d} u_\gamma^\prime }{\mathrm{d} t^\prime} \right]_\mrm{exp} \, ,
\end{equation}
here $\left[\mathrm{d} u_\gamma^\prime / \mathrm{d} t^\prime \right]_\mrm{exp}$ represents the photon cooling term due to the expansion of the plasma.

Second, we calculate the evolution of non-thermal leptons, muons, pions, protons and neutrinos from Eq.~\ref{eq:transport} relying on  a modified version of the \textsc{AM$^3$} code \citep{Gao:2016uld}. The latter has  been altered with respect to its original version in the following aspects:
\begin{enumerate}
    \item Bethe-Heitler pair-production. Photo-hadronic pair production is implemented following \citet{Kelner:2008ke}; see also  \citet{Cerruti:2021hah, Rudolph:2022ppp}. 
    \item Plasma expansion. The adiabatic expansion of the plasma yields a cooling term $\dot{x} =2 \beta \Gamma c E / 3 r$ for all charged particles. Dilution due to volume expansion is treated as an effective escape term  of $\alpha (x) = 2 \beta \Gamma c / r$~\citep{Vurm:2008ue}. For protons and electrons, the escape term is modified such that their evolution matches one of the thermal populations defined through $\rho^\prime(r)$. 
    \item Proton-proton interactions. The large plasma density enables efficient proton-proton interactions. These are newly implemented, and we use the pion multiplicities provided in \cite{Kelner:2006tc} and the updated cross-section from \cite{Kafexhiu:2014cua}; the decay processes of pions and muons are implemented as in \citet{Lipari:2007su}, while the kaon contribution is neglected for simplicity.
    \item Quantum synchrotron radiation. Owing to the large magnetic field  [$\mathcal{O}(10^8\, \mrm{G})$], quantum synchrotron radiation introduces a cut-off in the electron synchrotron emission that may be expressed as a modified synchrotron emission kernel \citep{Brainerd:1987}. For a more efficient computational implementation, we follow \citet{Imamura:1985} and model this effect as a simple upper limit on the emitted photon energy (or a lower limit on the energy of the radiating electron). We have however verified that the results are in good agreement with respect to the ones obtained through the full treatment proposed in \citet{Brainerd:1987}.
    \item Cooling of intermediate particles. Relying on the \textsc{AM$^3$} code version used in \citet{Rudolph:2022ppp}, we include  adiabatic, synchrotron, and inverse Compton cooling of intermediate pions and muons. Note that  we neglect any contribution from kaons, following   the procedure outlined in \citet{Hummer:2010vx}. 
\end{enumerate}

Finally, photons are evolved with a separate, novel code that contains both emission/absorption from non-thermal and thermal particles,  accounting for Comptonization due to the thermal electron population. The emission and absorption terms due to non-thermal processes are calculated with \textsc{AM$^3$} and passed to the photon simulation code. In addition, we account for emission and absorption due to thermal bremsstrahlung, cyclotron and double Compton emission \citep{1986:RybickiLightman, Vurm:2012be, Begue:2014kxa}. As the electrons are in thermal equilibrium, Kirchhoff's law can be applied. As a consequence,  the emission $\epsilon (x)$ and absorption rate $\alpha(x)$ are linked through the following relation
\begin{equation}
 \epsilon (x) = \frac{\alpha(x) }{e^{x/\theta_e} -1} \, , 
\end{equation}
where $x$ represents the momentum of the photon and $\theta_e = k_B T_e / m_e c^2 $ is the dimensionless electron temperature. \\
Comptonization by a Maxwellian distribution of electrons is included through the Kompaneets equation: 
\begin{equation}
    \frac{\partial n}{\partial t} = \frac{\sigma_\mrm{T} n_\mrm{e} c}{x^2}\frac{\partial}{\partial x} x^4 \left( \theta_e \frac{\partial n}{\partial x} + n^2 + n \right) \, .
    \label{eq:kompaneets}
\end{equation}
Here $n$ is the photon occupation number, $n_e$ the total number density of thermal electrons and $\sigma_T$ is the Thomson cross-section. 
Adiabatic expansion leads to the same cooling and escape terms introduced above. 
The evolution of the photon population in energy and time is calculated relying on the Chang and Cooper scheme \citep{1970:ChangCooper, 1996:ParkPetrosian}, where we choose an equally spaced grid in $ln(x)$. Note that for all particle species we omit escape terms, since we are still in the optically thick regime even at the largest radius of $10^{12}$~cm.

\end{document}